# $KTaO_3$(001) Preparation Methods in Vacuum: Effects on Surface Stoichiometry, Crystallography and in-gap States

Running title: $KTaO_3$ Surface Preparation Methods in Vacuum

Running Authors: Lucero Manzano et al.

Andrea M. Lucero Manzano [1,2,3, a)], Esteban D. Cantero [1,2], Emanuel A. Martínez [4,5], F. Y. Bruno [5], Esteban A. Sánchez [1,2], and Oscar Grizzi [1,2]

[1] Centro Atómico Bariloche, Comisión Nacional de Energía Atómica (CNEA), Bariloche 8400, Argentina
[2] Instituto de Nanociencia y Nanotecnología, Consejo Nacional de Investigaciones Científicas y Técnicas, (CONICET), Bariloche 8400, Argentina
[3] Universidad Nacional de Cuyo, Facultad de Ciencias Exactas y Naturales, Mendoza 5500, Argentina
[4] Dipartimento di Scienze Fisiche e Chimiche, Università degli Studi dell'Aquila, L'Aquila 67100, Italy
[5] GFMC, Departamento de Física de Materiales, Universidad Complutense de Madrid, Madrid 28040, Spain

[a)] **Electronic mail:** andrealucero@cnea.gob.ar

**Abstract**

$KTaO_3$ single crystals with different orientations are used as substrates for the epitaxial growth of thin films and/or as hosts for two-dimensional electron gases. Due to the polar nature of the $KTaO_3$(001) surface, one can expect difficulties and challenges to arise in its preparation. Maintaining good insulating characteristics without adding undesirable in-gap electronic states, obtaining good crystalline order up to the top surface layer, a sufficiently flat surface, and complete cleanliness of the surface (without water, C or OH contaminants), are in general difficult conditions to accomplish simultaneously.

Cleaving in vacuum is likely the best option for obtaining a clean surface. However, since $KTaO_3$ is cubic and lacks a well-defined cleavage plane, this method is not suitable for sample growth or reproducible device fabrication. Here, we systematically evaluate the effect of typical preparation methods applied on the surfaces of $KTaO_3$(001) single crystals. In particular, we used annealing in vacuum at different temperatures, light sputtering with $Ar^+$ ions at low energy (500 eV) followed by annealing, heavy $Ar^+$ ion bombardment and annealing, and grazing $Ar^+$ ion bombardment under continuous azimuthal rotation combined with both annealing in vacuum and in $O_2$ atmosphere. Possible side effects after each treatment are evaluated by a combination of techniques, including low-energy ion scattering at forward angles, Auger electron spectroscopy, low-energy electron energy loss, X-ray photoelectron spectroscopy, low-energy electron diffraction, and time of flight-secondary ion mass spectrometry. Advantages and shortcomings of each preparation method are discussed in detail.

# I. INTRODUCTION

Research on two-dimensional electron gases (2DEG) in oxide heterostructures began two decades ago.[1,2] Since then, many of their electronic properties have drawn considerable attention from the scientific community. In particular, interesting phenomena arising from the confinement of electrons to two-dimensional motion have been explored such as their high electronic mobility and strong electron interactions, among others.[3-9] This led to a wide range of applications in quantum technologies, spintronics, and oxide electronics.[10-13] Pioneering studies in this field began with the realization of 2DEG in the $LaAlO_3/SrTiO_3$ (LAO/STO) system,[1, 14, 15] and later expanded to heterostructures with other transition metal oxide (TMO) materials, including $LaTiO_3$ (LTO) and $KTaO_3$ (KTO). [16-20] A critical breakthrough in achieving high-mobility 2DEGs at the renowned LAO/STO interface was the development of a surface preparation method to obtain $TiO_2$-terminated STO substrates, emphasizing the crucial role of surface preparation.[21-22]

Progress in achieving similar results on the $KTaO_3$(001) surface is underway, but many unknowns remain, particularly regarding the chemistry of adsorbed species before sample growth.[23-25] The use of KTO substrates to form 2DEGs presents some particular characteristics that have attracted a significant interest in recent years.[26] These include its high relative permittivity, excellent crystallographic quality, high electron mobility, and compatibility for constructing heterostructures with other oxides. [12, 17, 19, 20, 27-29] Recent studies have revealed that the $KTaO_3$(001) surface hosts rich and diverse physics, including long-range magnetic order,[30] Rashba spin splitting,[31] polarity compensation mechanisms,[32] and charge-to-spin conversion.[33] Thus, continued exploration of surface preparation methods for this orientation is both timely and essential to enable future advances. Among the various methods currently being applied for the formation of 2DEGs on KTO substrates, some involve preparing the surface by bombarding it with ions (sputtering) or growing one or more specific layers on it, *e.g.*, surface engineering to create oxygen vacancies ($V_O$), or depositing a capping as a protection layer for devices development,[10, 34-37] which require the achievement of pure, namely impurities-free surfaces along with good crystallographic order. Despite the excellent quality recently demonstrated in epitaxial $KTaO_3$ thin films, most experimental studies still rely on commercially available $KTaO_3$ single crystals; therefore, surface preparation methods remain highly relevant.[38, 39]

The systematic and detailed investigation of these preparation methods is crucial for advancing the possibility of forming 2DEGs directly on KTO substrates, by inducing and stabilizing a charge accumulation that produces a polar discontinuity on its surface, promoting the formation of a conductive electron layer through chemical doping or the formation of oxygen vacancies. [26, 32, 41] As in many insulating compounds (particularly in oxides), surface preparation presents a challenge because it requires a trade-off between preserving the insulating character of the substrate and simultaneously obtaining a sufficiently clean surface, which is even more difficult on highly reactive polar surfaces.[23, 42-45] In the case of KTO, this issue is of high importance because the various crystalline surfaces on which 2DEG fabrication has been studied exhibit a varied degree of polarization.[26, 32, 46, 47] The contamination of KTO surfaces with water or OH is often difficult to quantify during the growth stages due to the low (or null) detection sensitivity of H in typical sample preparation equipment, and the natural presence of oxygen in the substrates, which makes hard (or even impossible) to distinguish it from the contribution of oxygen impurities due to surface contamination. Part of these analyses can be performed in studies that combine several surface science experimental techniques.

In a recent work,[37] we have demonstrated that the use of Time-of-Flight Secondary Ion Mass Spectrometry (TOF-SIMS) depth profiles with very low-energy sputtering allows for determining fundamentally important information about KTO-based 2DEG systems, regarding the composition of the different layers and their respective interfaces, as well as the distribution of oxygen vacancies in these types of samples. This information is of significant interest to the oxide materials community; however, it is typically very difficult to quantify experimentally.

The motivation of this work is to explore the ability to reveal the intricacies and side effects related to the different surface preparation methods by combining TOF-SIMS studies with other qualitative surface analysis techniques (Direct Recoil Spectrometry with Time-of-Flight analysis (TOF-DRS), Auger Electron Spectroscopy (AES), Low Energy Electron Spectroscopy (EELS), and X-ray Photoelectron Spectroscopy (XPS)), to elucidate which surface cleaning method is most appropriate for the primary goal of formation of a 2DEG on KTO or for eventual other uses of this substrate. In particular, we analyze effects due to sample annealing (commonly used to remove surface contamination from H) and low-energy (~keV) Ar ion sputtering.

## II. Samples and methods:

The samples were $KTaO_3$(001) single crystals obtained from several manufacturers (PI-KEM, MaTeck, SurfaceNet), with 5 mm x 5 mm x 0.5 mm dimensions, initially transparent and highly polished by the manufacturer. In some cases cleavage in air was used in order to compare with fresher surfaces having no previous treatment.

Annealing was done by electron bombardment with a tungsten filament mounted below the sample mounting plate (without exposing the sample directly to the electron beam). The temperature was measured with a thermocouple mounted near the sample. The thermocouple readings were previously calibrated with another thermocouple mounted directly on top of the surface (used only for this calibration and then removed for the experiments). Most of the sputtering with Ar ions was done using a low-energy gun, typically in the range of 0.5 to 1.5 keV, mounted at an incidence angle of 45$^o$. Another method of sputtering consisted in grazing ion bombardment with 10-20 keV $Ar^+$ (typically between 2 and 4$^o$ with respect to the surface plane) while constantly rotating the azimuthal direction of the sample. This method produces very flat and clean surfaces in metals and insulators, and typically generates a lower damage than sputtering at larger incidence angles with respect to the surface plane. [48]

TOF-DRS, LEED, AES, and EELS were conducted in a custom-built UHV chamber, detailed in Ref.[49], which is linked to a low-energy ion accelerator (2–100 keV) that operates in pulsed mode. TOF-DRS experiments utilized 5.5 keV $Ar^+$ ions, with recoiled and scattered ions, and neutral atoms detected at the end of a 1.73 m long drift tube positioned at a 45° scattering angle (forward scattering). The same chamber contains a LEED system with channelplate amplification, as well as standard and monochromated electron guns for AES and EELS, respectively, along with a He lamp for UPS. A spherical electron energy analyzer is mounted on a rotating platform to facilitate angle-resolved measurements, and in all cases, it operated in constant retard ratio mode.

XPS measurements were carried out in a SPECS instrument using an Al Kα monochromatized X-ray source (1486.6 eV) with a Phoibos 150 electron analyzer working in constant pass energy mode, at room temperature and pressure better than $10^{-9}$ mbar.

TOF-SIMS depth profiling experiments were performed using a TOF.SIMS 5-100, IONTOF GmbH, installed at GIA-CNEA, Argentina, in dual beam mode. A pulsed $Bi^+$ beam was used for analysis in Spectrometry mode (bunched mode), with a kinetic energy of 30 keV and a pulsed current of 1 pA, rastered randomly on a 100 × 100 μ$m^2$ area with 128 × 128 pixels. Most of the sputtering was performed with $Cs^+$ ions with a kinetic energy of 250 eV and a current of 8 nA. In a few cases, we used 500 eV $Cs^+$ (with a current of 38 nA) for sputtering. The sputtered area (250×250 μ$m^2$) was larger than the analysis area in order to avoid crater edge effects during SIMS analysis. Non-interlaced mode was used. Charge compensation was done with a low-energy electron source. The internal pressure of the TOF-SIMS chamber was maintained in ~$9\times10^{-10}$ mbar. To convert the sputter time to a depth scale, the depth of the sputtered craters was measured with both stylus and optical profilometers, and an average sputter rate was obtained.

# III. RESULTS AND DISCUSSION

## *A. Annealing effects:*

One of the most accepted methods to prepare the surface of KTO crystals for creating 2DEGs is annealing in ultra-high-vacuum (UHV) at temperatures around 800 K and, in some cases, up to higher temperatures.[35, 36] This method is commonly used before the deposition of a thin reducing layer (Al, EuO, $LaAlO_3$, etc) that is responsible for draining oxygen near the KTO surface, thus creating the $V_O$ necessary for developing the 2DEG. [37, 50-52] Here, in order to follow the evolution of the surface versus temperature we use TOF-DRS, which is sensitive to H, extremely sensitive to the top composition, and can eventually delineate between bulk and surface-adsorbed oxygen[48] and produces negligible damage. In this case, we used polished surfaces, directly provided by the manufacturers, without any previous treatment.  Figure 1a shows TOF-DRS spectra for the KTO surface taken with 5.5 keV Ar+ ions after annealing the surface at the specified temperature. Below 600 K the spectra are dominated by O and H recoil peaks coming from OH groups and water adsorbed on the surface. Above 700 K, the H and O recoil peaks decrease and a new recoil peak in the C position appears while at the same time the scattering peaks from the substrate begins to emerge. This indicates that OH or water desorbs in this range of temperatures, while a thin C layer remains adsorbed. With increasing temperatures, above 900K structures associated to Ta and K start to appear, indicating that the C layer thickness is reduced below a full monolayer, otherwise the substrate recoiling peaks would not be detected. Further annealing or annealing at higher temperatures does not remove this C layer. This behavior was systematically observed in the three samples tested. The surface crystallography followed by LEED, however, depended on the sample: the sample from MaTeck presented a clear 1x1 pattern after annealing (shown in Figure S1 of the supplementary material), while the other samples presented a less ordered surface with spots that depended on the surface position, being more typical of faceted surfaces. Concerning the bulk order, the Laue diffraction patterns showed single-domain crystal with very sharp spots (inset in Figure S1 of supplementary material).

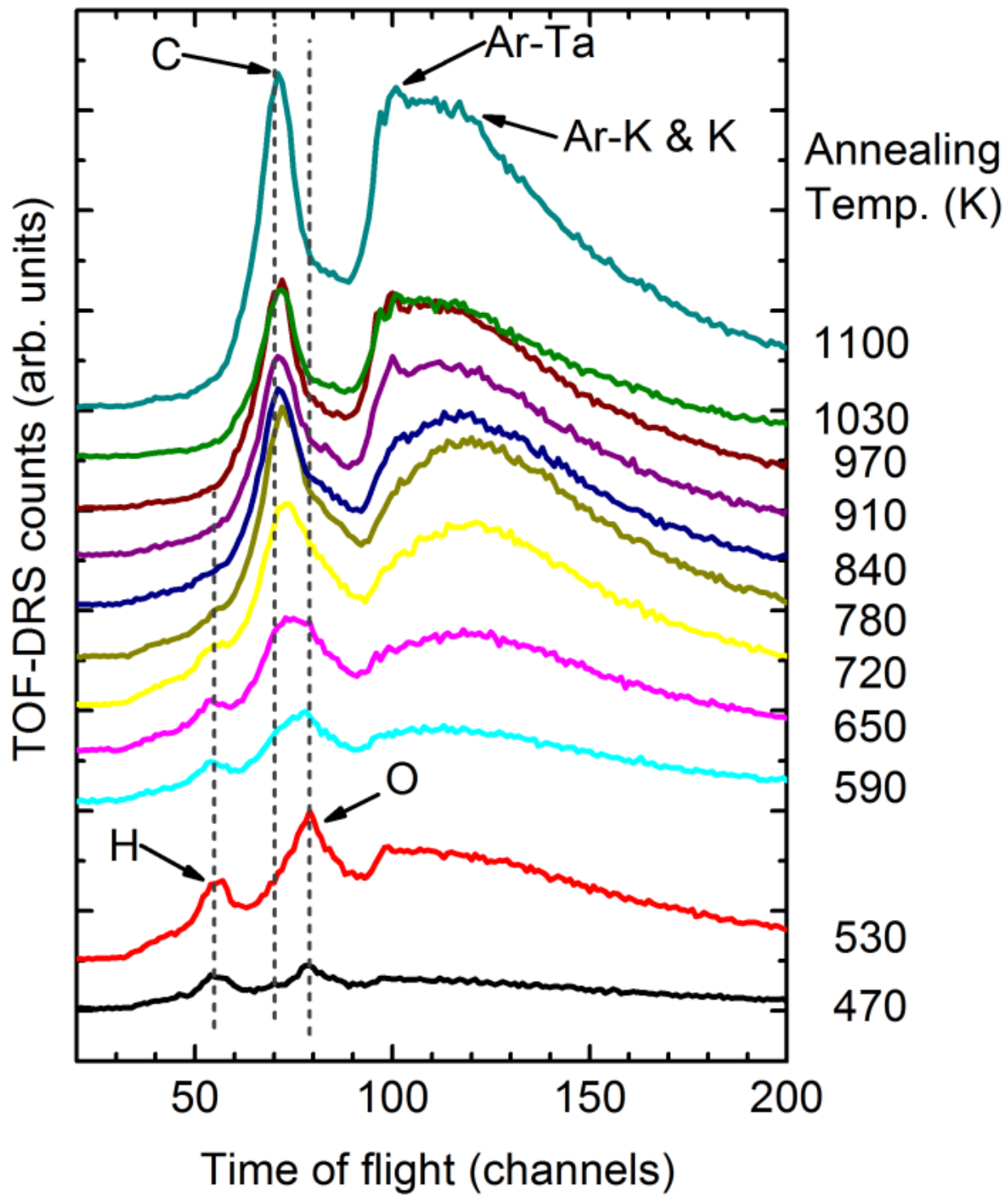

FIG. 1. TOF-DRS spectra measured at 45° scattering angle with 5.5 keV $Ar^+$ impinging at 20° incidence, taken as a function of the sample temperature.

When the surface cools down, we observe a strong reaction to residual water vapor present in the vacuum system, similar to what we have observed previously for another polar surface (ZnO(0001)),[53] i.e., the H and O recoil peaks increase significantly in 15-30 min, even in a vacuum better than $1x10^{-9}$ mbar obtained by the combination of turbomolecular and ionic pumps. The behavior described above validates the procedure used in the preparation of KTO samples previous to Al deposition,[37] which consisted of annealing the sample at 780 K, and then keeping the sample at temperatures around 600K during Al deposition. This process removes most of the contamination and preserves the surface free of water or OH deposition during layer growth. It should be noted that systems prepared without this annealing procedure did not form a 2DEG.[37]

Potassium is often mobile or volatile, tending to migrate in K compounds when they are heated. To quantitatively check the K content near the KTO surface during annealing we used Auger electron spectroscopy (AES), which can detect K below the thin C layer. In this case, we cleaved a KTO sample in air and immediately introduced it in the UHV chamber. Then we proceeded to anneal the sample with the cleaved surface exposed to the AES electron beam. This surface showed a rather good 1x1 LEED pattern already before heating (Figure S1 of supplementary material) and this pattern improved during annealing. The set of AES spectra is shown in Figure 2a. Due to the size of the cleaved surface (0.5 mm x 5 mm), part of the C and O AES signal may come from the sample holder, since the spot of the electron beam is slightly larger than 0.5 mm. The observed K signal only comes from the sample; its dependence on temperature is shown in Figure 2b. We observe that K already starts to decrease above 600 K and its dependence becomes much more pronounced after 800K. This behavior shows that annealing is necessary to remove the OH/water layer but care must be taken to avoid large losses of K.

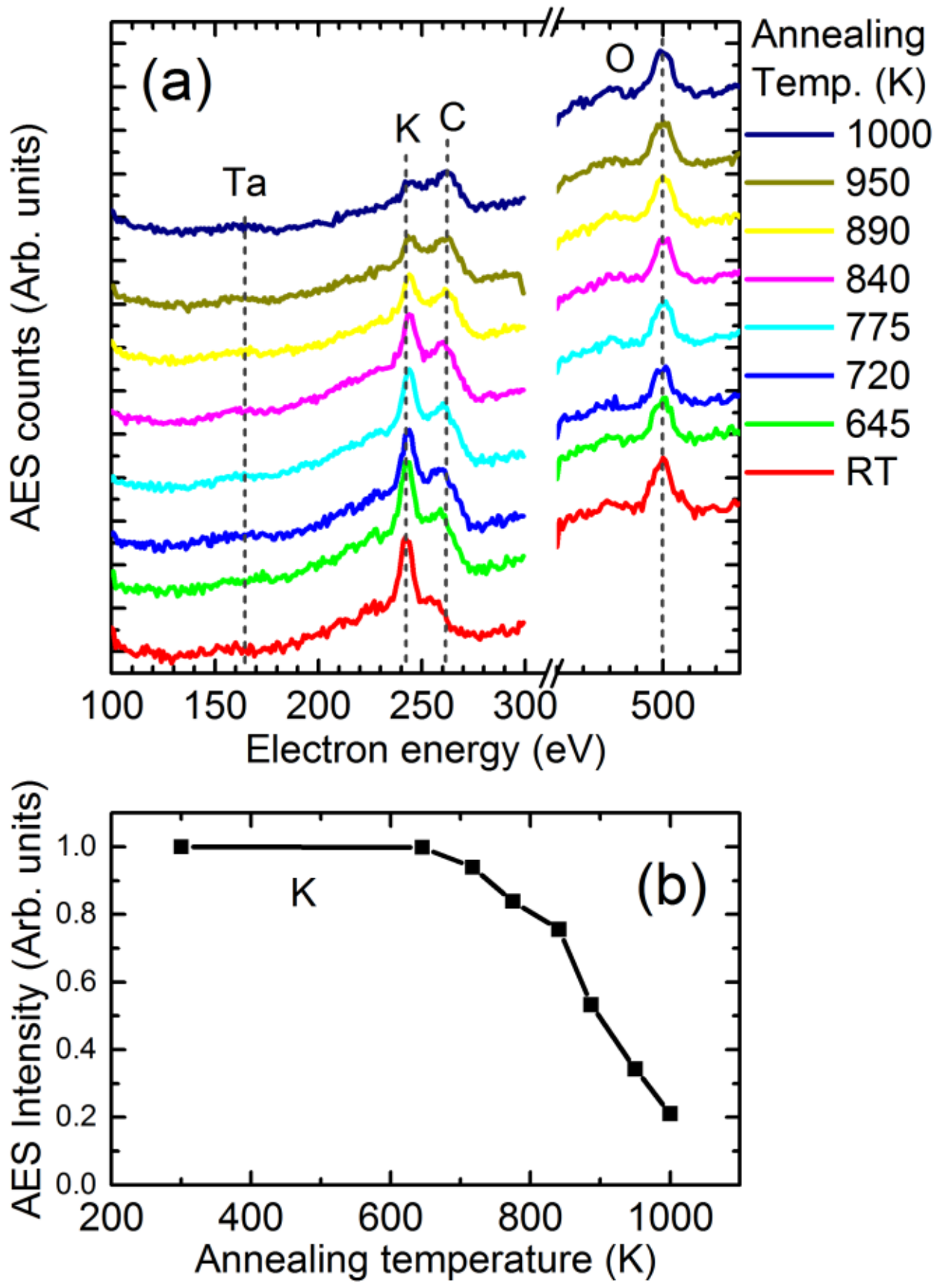

FIG. 2. (a) AES spectra for a cleaved "in air" $KTaO_3(001)$ surface after annealing at the indicated temperature. (b) Potassium intensity obtained as K peak integral versus annealing temperature.

## *B. Effect of cycles of annealing and sputtering*

If the KTO surface were used as an ordered substrate for epitaxial growth it might be desirable to fully remove the thin C layer that remains after annealing. In order to attempt this, we first annealed the surface in an $O_2$ atmosphere (typically in $10^{-6}$ mbar of $O_2$) to check whether C would combine with O and leave the surface as CO, but this method was unsuccessful as almost no change in the C content was found (we did not try atomic oxygen). We then used different cycles of $Ar^+$ sputtering combined with annealing at temperatures below 850 K. In particular, we performed sputtering at low energy (0.5-1.5 keV) and large incidence angles (around 45º) or higher sputtering energy (in the range of 10-20 keV), but low incidence angles (2-4º), always combined with annealing in UHV or in $O_2$ atmosphere, and continuous rotation of the azimuthal angle. Understanding the effects of $Ar^+$ ion bombardment is relevant since it was also proposed as a method to generate oxygen vacancies and produce the 2DEG.[34, 54] In this section, we describe the effect of these sputtering methods combined with annealing of the KTO surface, which complement the previous studies.[54, 55]

Figure 3a shows the evolution of the AES spectra of the KTO surface (from MaTeck sample) with an increasing number of sputtering cycles using 0.5 keV $Ar^+$ ions, corresponding to a dose of approximately $10^{15}$ $Ar^+$ ions/$cm^2$ per cycle. After each cycle the surface was annealed in the range of 700 -800 K. We observe a clear decrease in the C content, together with an increase in the substrate components signals, when compared with the initial spectrum. After the fifth cycle, Ta and O remain stable (or experiment a small increase) while K clearly starts to decrease, even when the annealing temperature was kept relatively low.

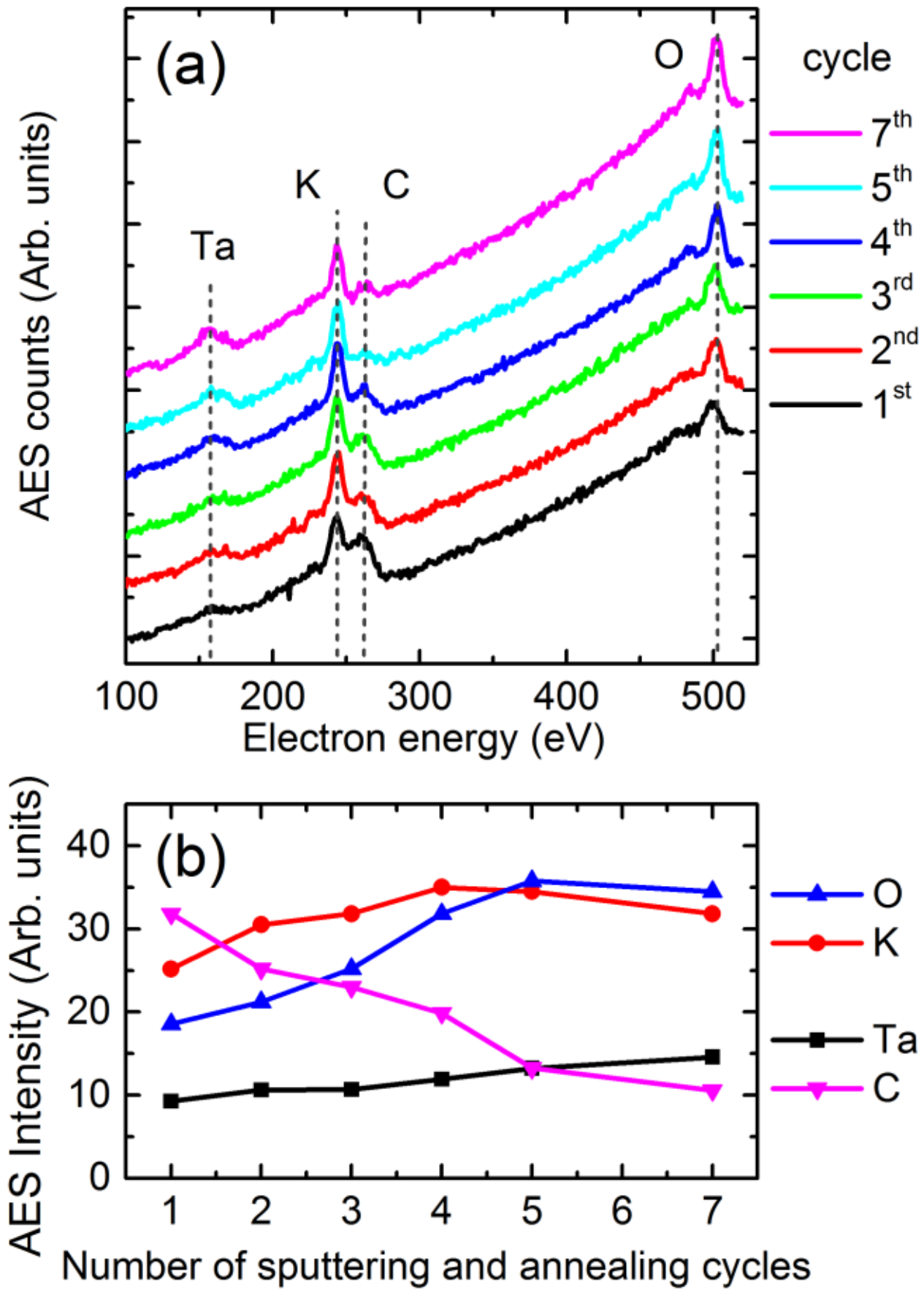

FIG. 3. (a) AES spectra vs. number of cycles of sputtering at 500 eV and annealing. (b) Evolution of AES intensity of relevant peaks.

We observed that, similar to ZnO(0001), as the surface becomes cleaner from C, it also becomes more reactive to water or OH adsorption. This is shown in Figure 4a by the green spectrum, taken several minutes after (*i.e.* during cool down) performing three sputtering cycles, note the H build-up. Further sputtering and annealing cycles produce the best (more representative) TOF-DRS spectrum for this surface, *i.e.* a spectrum with O, Ta, and K components, free of H and or C components. The LEED patterns initially improved during the first few sputtering and annealing cycles, but with an increasing number of cycles, the sharpness started to get worse, and it was not possible to recover it by changing the annealing temperature, time, or atmosphere (using $O_2$). This proves that, due to the multicomponent nature of $KTaO_3$, annealing is not sufficient to recover crystallinity. Sputtering at high energy (10-20 keV) and low incidence angles (2-4$^{o}$), coupled with constant azimuthal rotation, and annealing after the sputtering cycle also was efficient to remove C and water (Figure S2 of supplementary material). However, it did not result in a clear improvement in the LEED pattern, as usually happens at metal and semiconductor surfaces.[43] Summarizing this section, we show in Figure 4b: 1) the AES spectrum for a KTO surface as received ("Pristine"), where the Ta peak is hard to see, 2) the AES spectrum after annealing ("A @ 850 K"), showing the three substrate components (Ta, K and O) and an apparent increase in C, which is attributed to removal of water or OH at the surface top, and 3) the spectrum for the KTO surface cleaned by grazing Ar sputtering and annealing ("GS+A"), without traces of C, but having a too large Ta peak. This increased content of Ta upon sputtering was described before[55] and will be further analyzed in the section below.

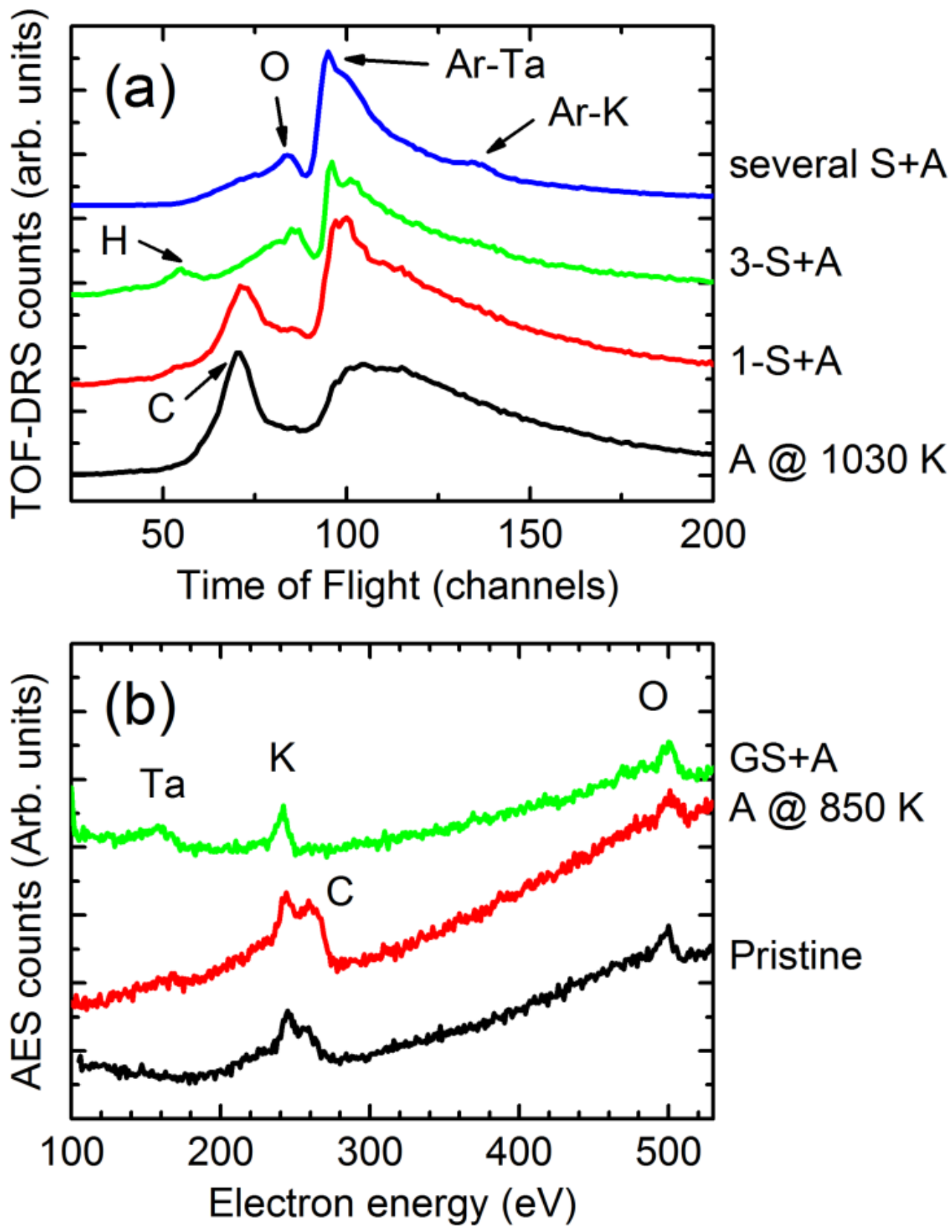

FIG. 4. (a) TOF-DRS spectra showing the effect of a number “n” of cycles of sputtering at low energy combined with annealing (n-S+A). The black spectrum corresponds to a sample annealed at 1030 K (A @ 1030K), before sputtering. (b) AES spectra at different surface preparation conditions: pristine or as received sample (Pristine), after annealing at 850 K without sputtering (A @ 850 K)), and after grazing $Ar^+$ ion bombardment and annealing (GS+A).

The effects described above show that preparation of the KTO surface implies a compromise, particularly regarding the top surface C content. In fact, some C content was already detected in the bulk by both XPS and TOF-SIMS profiling, suggesting that C could be present as an impurity. The amount of C in the bulk was similar for the samples from different manufacturers, reaching values of 1-3 % according to XPS (Figure S3 of supplementary material). All sputtering methods, varying energy, time, and incidence angle were efficient to remove the C content in the surface, but also generated side effects. If the KTO surface will be used as a substrate for epitaxial growth, then, probably some sputtering to reduce C will be the best option, however, if the surface will be used to assemble a 2DEG, where the presence of uncontrolled states may ruin the device, sputtering may not be an option.

As it has been discussed previously, $Ar^+$ bombardment also has a strong effect on the work function and the band gap.[54] The states generated in the band gap by $Ar^+$ bombardment were studied by XPS[55] and by optical spectroscopy.[26, 54] Additional evidence about the generation of states within the gap by $Ar^+$ bombardment can be obtained by EELS, as we have shown before for the $TiO_2$ surface.[56] In Figure 5 we show an EELS spectrum for a surface as received, before annealing or any other treatment, and the corresponding spectrum for a sample after sputtering cycles at 500 eV plus annealing. If the sample were a perfect insulator, there should be no accessible electronic states within the gap. This implies that the energy spectrum produced by electron bombardment in EELS should not show electrons with energy losses in the area indicated by the "energy gap" legend in Figure 5. The fact that electrons with energies in this area appear is related to the possibility of exciting electrons that have energies lower than the gap, which would indicate the presence of electronic states within it as reported for example in Ref.[26]. Clearly, despite the presence of some C and water/OH, the pristine sample has a much better-defined gap than the “clean” or “prepared” surface. Similar information about the changes developed under the different treatments can be obtained by XPS, as is shown in Figure S3 of the supplementary material, and elsewhere.[55]

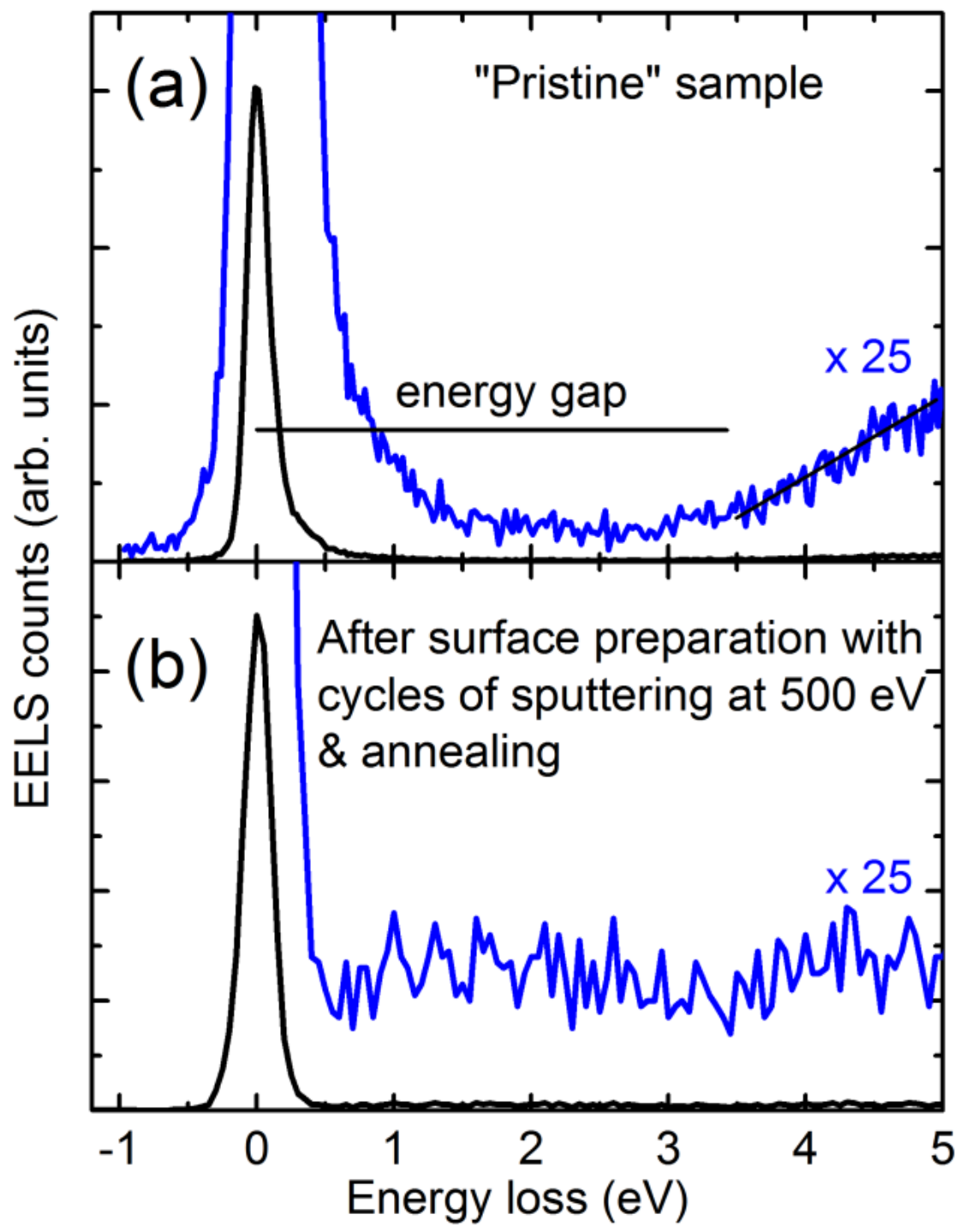

FIG. 5. (a) EELS spectra for $KTaO_3$(001) sample previous to any surface treatment, *i.e.*, “pristine” or “as received” and (b) for a sample prepared by cycles of sputtering with $Ar^+$ ions at 500 eV and annealing.

## *C. Surface preparation effects on deeper substrate layers as seen by TOF-SIMS*

TOF-SIMS combined with Cs bombardment at low energy (250-500 eV) allows measuring the variation of the intensity as a function of depth for some ions of interest (depth profiles). This method is useful for assessing the depth extent of the effects inside the substrate due to the different surface treatments, *i.e.* due to sputtering and annealing or only annealing of the KTO sample. Here, the low intensity of the Bi beam used for analysis produces negligible damage, and the use of a heavy ion (Cs) at low energy (typically 250 eV, in some cases 500 eV) for profiling modifies only a very shallow region, thus conserving the major effects of the previous surface treatments. To corroborate this assumption in all cases the spectra for the treated surfaces are compared with equivalent spectra taken at pristine surfaces (without any treatment). In particular, in this section, we discuss the depth profiles for: 1) a sample exposed to heavy Ar bombardment (1 -2 keV) and annealing (which we refer as “S + A KTO”), 2) a sample that was annealed in vacuum at 780 K belonging to a batch of samples used for 2DEG device fabrication (“Anneal KTO”), and 3) a pristine sample mounted as received without any treatment (“pristine”). The treatment of the samples was done *ex-situ, i.e.*, they were prepared in a different vacuum chamber, and then they remained in air for long periods of time (of the order of a month) and finally mounted in the TOF-SIMS instrument for analysis. Because of this exposure to air, the initial part of the profiles has a non-controlled contamination; nevertheless, we will show that the main effects of the different surface treatments persist. The “Anneal KTO” sample has a capping layer of $Si_3N_4$, deposited *in-situ* immediately after the annealing treatment to preserve the KTO surface (interface in this case) from the air contamination.

Figure 6 shows some representative plots for K, Ta, and O ions as a function of sputter time for the different samples: negative ion profiles for “S+A KTO” (panel (a)), corresponding positive ion profiles (panel (b)), negative ion profiles for “Anneal KTO” (panel c)) and negative ion profiles for “pristine” samples. In cases where the more abundant ion tends to saturate, we used a less abundant isotope. In most cases, the profile was done with Cs at 250 eV, while in the case of the “Anneal KTO” sample it was done with Cs at 500 eV. In this case, the different Cs current and different sputtering yield produces a sampling rate that is almost ten times faster at 500 eV than that for Cs at 250 eV. This difference is evidenced in the lower horizontal scale (sputtering time). The upper horizontal scale is the depth in nm, obtained by calibration of crater depths done with profilometers after each measurement. As expected the pristine surface has initial variations, expanding up to 2-4 nm that are related to contamination due to the exposure to air. After this initial part is removed (sputtered away) all the signals remain essentially constant. In contrast, the “S+A KTO” sample presents variations in all the signals extending in some cases beyond 10 nm. The negative ion profiles are consistent with a depletion of K up to this depth and an enrichment of Ta in that zone. Here we recall that TOF-SIMS detects only ions, and the fraction of sputtered species that escape as ions may have strong variations due to “matrix” effects, *i.e.,* variations in the charge exchange processes taking place when the ion traverses the surface. To verify this, we show the positive ion profiles for the “S+A KTO” surface in panel b. We see that after the first 2-4 nm affected by surface contamination, $K^+$ remains lower than average and $Ta^+$ higher than average like the behavior of the negative ion, corroborating that these variations correspond to K depletion and Ta enrichment, and not SIMS matrix effects. The effects due to surface treatment are better represented in Figure 7 by 3D plots of the distribution of K and Ta. The variation of O profiles in the “S+A KTO” surface is less clear, since they show opposite variations for positive and negative ions in the important region between 2 and 10 nm, evidencing the presence of matrix effects. In fact, this variation could be related to the depletion of K in that region, since it is known that the presence of alkalis at surfaces (strong electron donors) tends to enhance the negative ion production of other adjacent elements. In contrast to the behavior of the “S+A KTO” surface, the surface that was only annealed (without sputtering) presents a very shallow or almost constant dependence with depth up to the capping interface for all the elements.

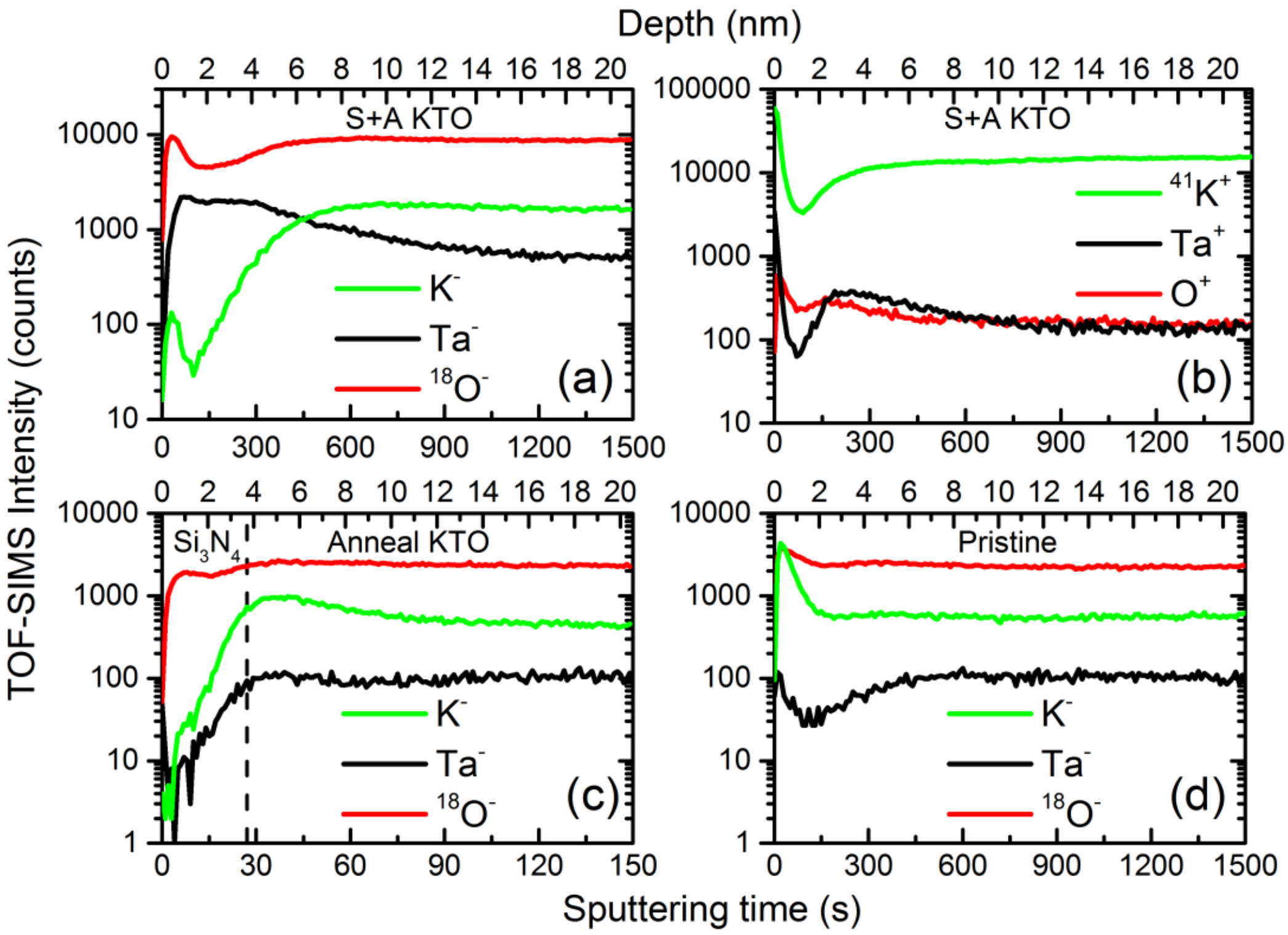

FIG. 6. TOF-SIMS Depth profiles of representative substrate ions (K, Ta and O) measured for the sputtered and annealed surface “S+A KTO” (a – negative ions, b – positive ions), for an annealed surface, without sputtering “Anneal KTO” (c – negative ions), and for a “pristine” or “as received” sample (d – negative ions).

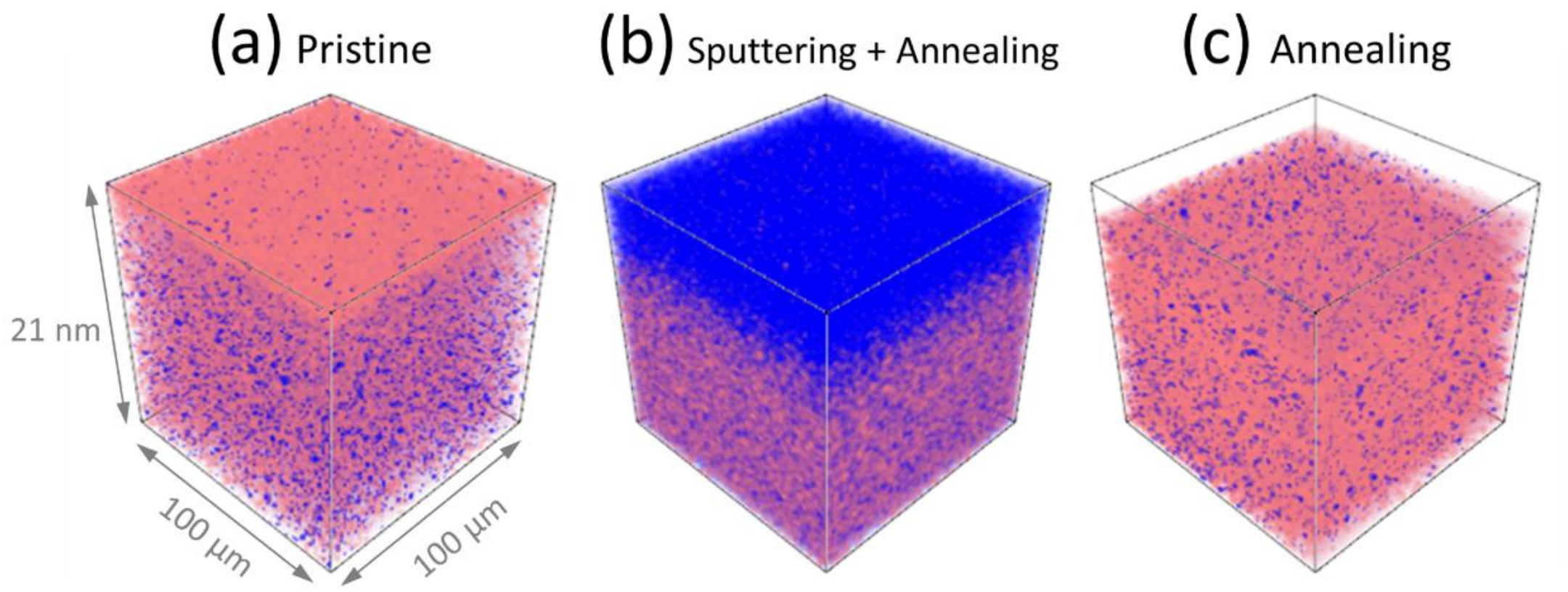

FIG. 7. 3D plots of the distribution of K (pink) and Ta (blue) in the following samples: (a) “pristine”, (b) “S+A KTO” and (c) “Anneal KTO”.

In a previous work we used sputtered clusters like $KTaO^-$ and $KTaO_3^-$ to obtain information about the $V_O$’s extent into the bulk generated by depositing a thin Al amorphous layer on a $KTaO_3$ surface.[37] In that work, we showed that these clusters presented opposite variations with depth and that the extension of the variations was correlated with the thickness of the Al layer. Here we show that these clusters are also sensitive to surface treatment. Figure 8a (Figure 8b) shows the profiles for $KTaO^-$ ($KTaO_3^-$) in the “S+A KTO” and in the “pristine” surfaces. As expected after the initial variation due to contaminants the clusters for the “pristine” surface remain constant, while those for the treated surface present a clear depletion in both clusters, which is consistent with the lack of K in that region. Similar profiles are observed for positive ions (Figure S4 in the supplementary material). Consistently with the individual elements dependence discussed above, these clusters show little variations with depth in the “anneal KTO” surface (Figure 8c).

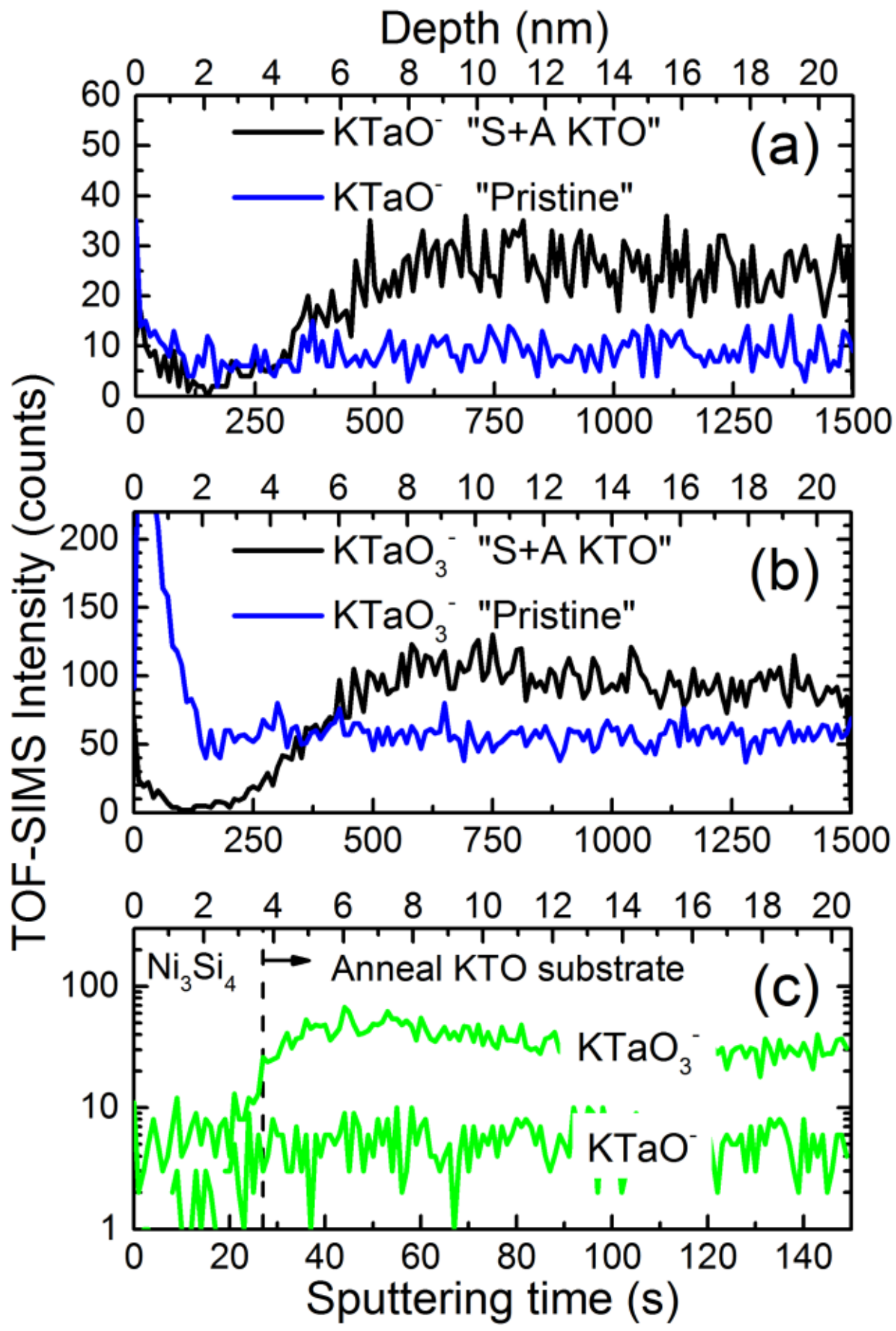

FIG. 8. Comparison KTO$^-$ (a) and $KTO_3^-$ (b) cluster TOF-SIMS profiles corresponding to the "S+A KTO" and "pristine" samples. (c) Same cluster profiles in the "anneal KTO" sample.

# IV. SUMMARY AND CONCLUSIONS

We studied the effects of several methods typically used to prepare the polar surface $KTaO_3$(001) for formation of 2DEGs or epitaxial sample growth. In particular, we tested annealing in UHV at different temperatures, and $Ar^+$ ion sputtering at different energies and incidence angle configurations combined with annealing. In order to measure the effects of the different methods we used a combination of surface science techniques having different sensitivity with respect to the depth, *i.e.*, TOF-DRS, AES, XPS, LEED, EELS, and TOF-SIMS. The results show that in contrast to what is observed in typical metal or semiconductor surfaces, in KTO a compromise must be assumed. In previous works[37] it was shown that without previous annealing, the method of producing 2DEG by depositing a thin film of Al does not work, resulting in an insulating Al/KTO interface, while after annealing a good quality 2DEG is formed. In the present work, we clearly show that the effect of this annealing is the elimination of most of the surface contaminant layer, particularly the OH/water adsorbed. Care must be taken to keep the surface at temperatures above 550K to avoid water and OH re-adsorption during all the steps of producing the 2DEG. Also the annealing temperature has to remain below 800 K to avoid dramatic losses of K. This method is the least disruptive of the surface electronic properties, however, it leaves some C on the top surface which is not removed even at higher temperatures. If this C content must be removed for any other application, we found that either low energy (500 eV) $Ar^+$ ion bombardment and annealing cycles or even better grazing $Ar^+$ ion bombardment at higher energies with continuous rotation of the azimuth, are efficient to remove this C layer, providing a well-ordered surface (as seen by LEED). This method, however, introduces additional states in the gap, clearly seen by EELS, and as a very small shoulder in the Ta 4f XPS peak, the latter coming from formation $Ta^{4+}$, and $Ta^{5+}$ states associated to $V_O$'s. Typical heavy $Ar^+$ ion bombardment (even when combined with annealing to reestablish crystalline order), often used to prepare metallic surfaces, will produce strong changes in the stoichiometry, particularly a depletion of K near the surface, which extends up to approximately 10 nm from the surface top.

## SUPPLEMENTARY MATERIAL

Additional results and figures are available in the Supplementary Material. In Figure S1 we present LEED patterns for KTO samples cleaned by cycles of sputtering and annealing, and for a cleaved sample. The inset shows a LAUE diffraction pattern. Figure S2 shows additional TOF-DRS spectra for KTO samples under different cleaning treatments. In Figure S3 we present XPS measurements for Ta 4f, K 2p, C 1s and the valence band for KTO samples after annealing and after cycles of sputtering and annealing. Finally, in Figure S4 we show a $KTaO^+$ TOF-SIMS positive ion cluster profile for a KTO sample after the sputtering plus annealing cycles.

## ACKNOWLEDGMENTS

This work has been supported by: grant PIP 2022-11220210100411CO funded by CONICET and grant 06/C025-T1 founded by UNCuyo. FYB acknowledges grant CNS2022-135485 funded by MCIN/AEI/ 10.13039/501100011033 and European Union NextGeneration EU/PRTR.

## AUTHOR DECLARATIONS

**Conflicts of Interest *(required)***

The authors have no conflicts to disclose.

**Author Contributions *(if applicable)***

A. M. Lucero Manzano: Formal analysis, Investigation, Validation, Writing – original draft, Writing – review & editing.

E. D. Cantero: Formal analysis, Investigation, Validation, Writing – original draft, Writing – review & editing.

E. A. Martínez: Resources, Validation, Writing – review & editing.

F. Y. Bruno: Conceptualization, Funding acquisition, Resources, Supervision, Validation, Writing – review & editing.

E. A. Sánchez: Formal analysis, Funding acquisition, Investigation, Supervision, Validation, Writing – review & editing.

O. Grizzi: Conceptualization, Formal analysis, Funding acquisition, Investigation, Supervision, Validation, Writing – original draft, Writing – review & editing.

## DATA AVAILABILITY

The data that support the findings of this study are available from the corresponding author upon reasonable request.